\documentclass[12pt]{article}

\usepackage{hyperref}
\usepackage{epsfig}
\usepackage{amsmath}
\usepackage{amsfonts}
\usepackage{amssymb}
\usepackage{eufrak}
\usepackage{mathrsfs}

\def\arXiv#1{\href{http://xxx.lanl.gov/hep-th/abs/#1}{#1}}

\def\be{\begin{equation}}
\def\ee{\end{equation}}
\def\ba{\begin{eqnarray}}
\def\ea{\end{eqnarray}}
\def\mc{\mathcal}

\def\appendix{{\section*{Appendix}}\let\appendix\section%
        {\setcounter{section}{0}
        \gdef\thesection{\Alph{section}}}\section}

\begin{document}

\thispagestyle{empty}
\def\thefootnote{\fnsymbol{footnote}}
\begin{flushright}
WIS/16/06-SEPT-DPP \\
 hep-th/0610008 \\
 \end{flushright}
\vskip 0.5cm

\begin{center}\LARGE
{\bf $2D$ Black Hole and Holographic Renormalization  Group}\\
\end{center}

\vskip 1.0cm

\begin{center}
\centerline{Satabhisa Dasgupta and Tathagata Dasgupta}

\medskip
\medskip
\centerline{{Department of Particle Physics,}} \centerline{{Weizmann
Institute of Science, Rehovot 76100, Israel}}
\medskip

\medskip
\centerline{\tt sdasgupt, tdasgupt@wisemail.weizmann.ac.il}

\vskip 0.5cm
\end{center}

\vskip 1.0cm

\noindent In \arXiv{hep-th/0311177}, the Large $N$ renormalization
group (RG) flows of a modified matrix quantum mechanics on a circle,
capable of capturing effects of nonsingets, were shown to have fixed
points with negative specific heat. The corresponding rescaling
equation of the compactified matter field with respect to the RG
scale, identified with the Liouville direction, is used to extract
the two dimensional Euclidean black hole metric at the new type of
fixed points. Interpreting the large $N$ RG flows as flow velocities
in holographic RG in two dimensions, the flow equation of the matter
field around the black hole fixed point is shown to be of the same
form as the radial evolution equation of the appropriate bulk scalar
coupled to $2D$ black hole.

\date{September 2006}

\vfill \setcounter{footnote}{0}
\def\thefootnote{\arabic{footnote}}
\newpage


\tableofcontents


\section{Introduction}
\setcounter{equation}{0}

In a theory with gravity, the subtle issue of scale dependence is
governed by the contribution of degenerate or pinched surfaces (for
spherical geometry). Considering a general noncritical string
background with graviton $g_{\mu \nu}=a^2(\phi) \hat g_{\mu \nu}(x)$
and dilaton $\Phi(\phi, x)$, the scale dependence roughly takes the
form \cite{pol-projects}

\be \beta_{\mu \nu} \frac{\delta S}{\delta g_{\mu \nu}}+\beta_{\Phi}
\frac{\delta S}{\delta \Phi}=\mbox{anomaly}=\frac{1}{2}
G^{IJ}\big(\phi^K(a)\big) \frac{\delta S}{\delta \phi^I}
\frac{\delta S}{\delta \phi^J}+...\,,
\label{scale-transformation-gravity}\ee where $\phi^I$s correspond
to different string states and $G^{IJ}$ represents the metric in the
space of those states, $S$ being some appropriate low energy
effective action for the boundary values of the fields at some
$\phi=\phi_0$. These boundary fields $g_{\mu \nu}(x,\phi_0)$ and
$\phi^I(x, \phi_0)$ are the initial data for the corresponding
critical background. The effective action $S$ is the value of the
classical action on the solution with these initial data. The
ellipses in the right hand side of the above expression represent
local function of the fields corresponding to the string states.
This anomaly arises when the world sheet completely collapses. The
first term in the right hand side is due to degenerate metrics
coming from the pinching of the world sheet at a given point
(assuming spherical topology). This is compensated by running of the
renormalized background fields to keep net scale invariance of the
effective action. This is reminiscent of the {\it Fischler-Susskind}
mechanism \cite{fs,fs1,clny1,clny2,clny3,pc}. The equation
(\ref{scale-transformation-gravity}) can be seen as Hamilton-Jacobi
equation for the classical action that captures the contribution of
the pinched spheres at least in one loop approximation.  This also
determines the sigma model beta functions for the boundary couplings
$\phi^I(x,\phi_0)$ \be \beta_{ws}(\phi)^I=a \frac{\partial
\phi^I}{\partial a}\,.\ee Thus the bulk evolution of the scalar
fields can be interpreted as the {\it holographic renormalization
group flows} for the boundary couplings. The running string tension
$a^2(\phi)$ has an interpretation of the physical scale for the {\it
holographic RG}.

This dynamics have been studied in the context of gravity/gauge
theory correspondence in $AdS_5 \times S^5$ geometry at large 't
Hooft coupling $\lambda=N g^2_{YM}$ \cite{dvv,vv,ev,db}, where the
classical evolution of the scalar fields $\phi^I$ coupled to $5D$
bulk can be interpreted as the renormalization group flows of the
$4D$ gauge theory couplings. The similar structural relationship
were argued to hold in the regime of small $\lambda$ too \cite{kv},
where supergravity is not a good approximation for the bulk
dynamics. Such a framework elevates the gravity/gauge theory
correspondence to open/closed string duality. However, it is not
obvious whether such a {\it holographic RG} exists in $2D$
noncritical background, which is an interesting question in the
context of open/closed string duality.

Earlier for the $2D$ case, we calculated the boundary flows as large
$N$ renormalization group flows of matrix quantum mechanics
\cite{dd-singlet}. Then extending the work in \cite{dd-singlet} we
studied the nonsinglet sector of $c=1$ matrix model by considering a
gauged matrix quantum mechanics on circle with an appropriate gauge
breaking term to incorporate the effect of world-sheet vortices
\cite{dd-nonsinglet}. A new coupling was introduced that would act
like vortex fugacity. The flow equations indicate
Berezinski-Kosterlitz-Thouless (BKT) phase transition around the
self-dual radius and the nontrivial fixed points of the flow exhibit
black hole like phases for a range of temperatures beyond the
self-dual point. One class of fixed points interpolate between $c=1$
for $R > 1$ and $c=0$ as $R \to 0$ via black hole phase that emerges
after the phase transition. The other two classes of nontrivial
fixed points also develop  black hole like behavior beyond $R=1$.
From a thermodynamic study of the free energy obtained from the
Callan-Symanzik equations shows that all these unstable phases do
have {\it negative specific heat}. The thermodynamic quantities
indicate that the system does undergo a first order phase transition
near the Hagedorn temperature, around which the new phase is formed,
and exhibits one loop finite energy correction to the Hagedorn
density of states. The fixed points with negative specific heat
suggest that this phase transition is associated with existence of
black hole in $2D$ gravity.

Motivated by the higher dimensional picture of holographic RG flow,
in this note we will give a natural interpretation of the
holographic RG in $2D$ by identifying the large $N$ renormalization
group flows in the matrix quantum mechanics, derived in
\cite{dd-singlet,dd-nonsinglet}, as the {\it radial evolution} of
the scalar fields coupled to the $2D$ bulk. Here the matrix quantum
mechanics acts as the boundary theory and the $2D$ bulk as the
holographic dual to the boundary theory. On one hand this examines
the validity of the large $N$ RG flows derived from the matrix
quantum mechanics and on the other hand illustrates an evidence of
holographic RG in the context of duality between matrix model and
$2D$ quantum gravity. Let us mention here that in large $N$ RG the
Callan-Symanzik equation for the physical obserbvables (which
basically gives the $2D$ nonperturbative background) is essentially
a world sheet hamiltonian constraint in the form of WdW equation.
When computed with all the $N^2$ quantum mechanical degrees of
freedom (that relates to having nonsinglets), this can accommodate a
much richer structure \cite{d-infl} than a semiclassical
minisuperspace WdW evolution of the background \cite{gm} and is
$\alpha'$ exact in computing the world sheet cosmological constant.
Whereas, the Callan-Symanzik equation for the holographic RG is a
classical Hamilton-Jacobi evolution, which on being linearized and
taken to the world sheet level, would reduce to minisuperspace WdW
evolution of the background with a leading order effect in
$\alpha'$.

Of particular interest is the flow equation for compactification
radius $R$ of the Euclidian time coordinate in matrix quantum
mechanics obtained in \cite{dd-nonsinglet}:

\be \frac{dR}{dl}=-h R\,, \ee where $h$ is some function of $R$ and
becomes large and positive as we approach the fixed points with
negative specific heat. It indicates a deformation of the target
space geometry if one identifies the RG scale in matrix quantum
mechanics to be the dilaton or the {\it radial direction} in the
holographic picture. Starting from the simple form of the flow
equation for $R$, near the fixed points with negative specific heat
suggesting black hole like phase of the flow equations of $c=1$
matrix model, one can derive the {\it cigar metric} of the $2D$
black hole \cite{EFR, MSW, Witten}. This is particularly nice as it
gives the indication that matrix quantum mechanics is capable of
extracting the $2D$ black hole metric that previously had been a
subject of continuous effort. It is important to note that the case
with $h = 0$ arises at the $c=1$ fixed point \cite{dd-singlet}. The
role of nonsinglet sector in matrix quantum mechanics is crucial in
obtaining black hole like behavior from the bulk background.

On the other hand, from the holographic RG point of view it can be
shown that the particular simple form of the $R$ trajectory can be
retrieved from the ratio of radial equations of motion of the bulk
scalar fields coupled to the $2D$ background. One serious
obstruction to such holographic picture for $2D$ cigar geometry is
the fact that the boundary RG equations derived from matrix quantum
mechanics are $\alpha'$ exact. The cigar background is more likely
to be visible in the high curvature regime where $\alpha'$ is
finite, while it is believed to be a Sine-Liouville background for
small $\alpha'$ \cite{FZZ-duality, KKK}. Thus unlike ADS/CFT
correspondence, using the dual supergravity description in the
holographic RG set up should no longer be useful to make contact
with the matrix quantum mechanics results and to see the cigar
metric. One then needs the more general framework of open/closed
duality to deal with the holographic RG for finite $\alpha'$ (see
for example \cite{kv}). However, in this paper we observe that the
$R$ trajectory determining the cigar metric being a ratio of flow
velocities of the bulk scalars, is independent of the curvature term
that contains all the $\alpha'$ dependence. It is thus consistent to
match the $R$ trajectory determined from the Holographic RG to that
derived from the $\alpha'$ exact computation of matrix quantum
mechanics.

The paper is organized as follows. In section $2$, we will briefly
review the large $N$ world sheet RG analysis leading to black hole
like fixed points and their thermodynamics. In particular we will
discuss the significance of the rescaling equation for the size of
the compactified space in capturing the change of target space
geometry around the fixed points. In section $3$ we calculate the
cigar metric of $2D$ black hole starting from the $R$ trajectory in
the large-$N$ in matrix quantum mechanics. In section $4$, via
Holographic RG picture, we interpret the origin of the $R$
trajectory in the boundary theory from the classical evolution of
bulk scalars.

\section{RG Analysis with Nonsinglet Sector}
\setcounter{equation}{0}

For a detailed account of the RG scheme see the original works in
\cite{dd-singlet}. A brief review of the work can be found in
section 3 of \cite{d-infl}. Here we will sketch the essential steps
one needs to add for analyzing RG flows involving nonsinglet sector
in MQM \cite{dd-nonsinglet} that result in appearance of fixed
points with negative specific heat. The concept of the RG scheme
arises from the interpretation of the very existence of the {\it
double scaling limit} as some kind of Wilsonian RG flow \cite{bz}.

In the double scaling limit, as the matrix coupling constant $g \to
g_c$, the average number of triangles in triangulations at any genus
$G$ diverges as

\be \langle n_G \rangle \sim (1-G)(\gamma_0-2)(1-g/g_c)^{-1} \,.\ee
Simultaneously with $N \to \infty$, the length of the triangles (the
regularized spacing of the random lattice)  $a\sim
N^{-\frac{1}{2-\gamma_0}}$ scales to zero to keep the physical area
$a^2 \langle n_G \rangle \sim N^{-\frac{2}{2-\gamma_0}} (1-g/g_c)$
or equivalently the string coupling $g_s \equiv N^2
(g-g_c)^{2-\gamma_0}$ fixed. The existence of double scaling limit
indicates that a change in the length scale induces flow in the
coupling constants of the theory in a way that one reaches the
continuum limit with correct scaling laws and the critical exponents
at the nontrivial IR fixed point determined by the flow equations.
The large $N$ world sheet RG in $MQM$ \cite{dd-singlet} is basically
the evolution of the two sets of parameters of the theory, the size
of the matrix $N$ and the cosmological constant (mapped into the
matrix coupling $g$) and all other matrix couplings, at the constant
long distance physics with the rescaling of the regularization
length in the triangulation of the world-sheet. In the Wilsonian
sense this is done by changing $N\to N+\delta N$ by integrating out
some of the matrix elements, which is like integrating over the
momentum shell $\Lambda-d\Lambda < |p| < \Lambda$, and compensating
it by enlarging the space of the coupling constants $g\to g+\delta
g$. Here the space of coupling constants contains both the matrix
coupling $g$ and the mass parameter $M^2$.

By integrating out a column and a row of an $(N+1)\times (N+1)$
matrix, the $c=1$ matrix partition function satisfies a discrete
relation

\be Z_{N+1}(g,M,R) = [\lambda(g,M,R)]^{N^2} Z_N(g',M',R')
\label{lambda-scaling} \ee with the following flow equations for the
couplings

\ba g' &=& g + \frac{1}{N}\beta_g(g,M,R)+O\Big(\frac{1}{N^2}\Big)\,,
\nonumber \\
M'^2 &=&
M^2+\frac{1}{N}\beta_{M^2}(g,M,R)+O\Big(\frac{1}{N^2}\Big)\,,
 \label{beta-defn} \ea and an auxiliary flow equation

\be \lambda(g,M,R)=1+N^{-1}~r(g,M,R)+O(N^{-2}) \,. \label{r-defn}
\ee Here $g', M'$ represent the renormalized couplings. Identifying
$1/N$ as the world sheet scale of the RG transformation, the
functions $\beta_g, \beta_M$ are interpreted to be the beta
functions for the corresponding couplings. The function $\lambda$
gives the change in the world sheet free energy (the string
partition function)

\be \mathcal{F}(N,g,M,R)=\frac{1}{N^2}\ln Z_N(g,M,R)
\label{string-partition} \,.\ee The auxiliary flow relation for
$\beta_{\lambda}$ is useful in studying thermodynamical
consequences. As $N$ is taken to be large, the world sheet scale
$1/N$ becomes infinitesimal. Then the discrete relation
(\ref{lambda-scaling}) for the Matrix partition function actually
translates into a continuous Callan-Symanzik equation satisfied by
the world sheet free energy with the function $r$ controlling its
inhomogeneous part

\be \Big[N\frac{\partial}{\partial
N}-\beta(g,M,R)\frac{\partial}{\partial
g}-\beta(g,M,R)\frac{\partial}{\partial M}+\gamma \Big]
\mathcal{F}(N,g,M,R)=r(g,M,R) \,. \label{C-S} \ee Here $\gamma$ acts
as the anomalous dimension. In \cite{dd-singlet} we have seen that
the beta functions $\beta_g, \beta_M$ computed by the large $N$ RG
are such that the homogenous part of the Callan-Symanzik equation
indeed determines the correct scaling exponents for $c=1$ matrix
model around the nontrivial fixed point. The inhomogeneous part is
related to some subtleties in the theory, like the {\it logarithmic
scaling violation} of the $c=1$ matrix model.

From the running of the prefactor of the partition function, written
in the renormalized couplings, analogous to the running due to the
wave function renormalization, the free energy is observed to change
sign near $R=1$ for small value of the critical coupling
\cite{dd-singlet}. This is reminiscent of the BKT transition at
self-dual radius triggered by the liberation of the world-sheet
vortices. The attempt in \cite{dd-nonsinglet} was to understand the
detail nature of the nontrivial fixed points of the flow that
describes the physics beyond this transition. To capture the effect
of vortices on the flows and the fixed points more clearly and to
introduce a new coupling that would act like vortex fugacity, in
\cite{dd-nonsinglet} we analyzed the behavior of the following
gauged matrix model with simple periodic boundary condition and with
{\it an appropriate gauge breaking term}

\ba &&\mathcal{Z}_{N}[g,\alpha,R]=\int_{\phi_{N}(2\pi
R)=\phi_{N}(0)}
\mathcal{D}^{(N)^2}A_{N}(t)~\mathcal{D}^{(N)^2}\phi_{N}(t)
\nonumber \\
&&\exp\Big[-(N)~\mbox{Tr} \int_0^{2\pi R} dt~
\Big\{\frac{1}{2}(D\phi_{N}(t))^2 + \frac{1}{2}\phi_{N}^2(t)
-\frac{g}{3}\phi_{N}^3(t)+\frac{A_{N}^2}{\alpha}\Big\}\Big] \,,
\nonumber \\
\label{AZ1N} \ea where the covariant derivative $D$ is defined with
respect to the pure gauge  $A(t) = \Omega(t)^\dagger\dot\Omega(t)$
by $D\phi = \partial_t \phi + [A,\phi]$, where $\Omega (t)\in U(N)$.
Expanding the covariant derivative, the partition function is
rewritten as \ba
\mathcal{Z}_{N+1}[g,\alpha,R]&=&\int_{\phi_{N+1}(2\pi
R)=\phi_{N+1}(0)}
\mathcal{D}^{(N+1)^2}A_{N+1}(t)~\mathcal{D}^{(N+1)^2}\phi_{N+1}(t)
\nonumber \\
&&\exp\Big[-(N+1) \mbox{Tr} \int_0^{2\pi R} dt~
\Big\{\frac{1}{2}\dot{\phi}_{N+1}(t)^2 + \frac{1}{2}\phi_{N+1}^2(t)
-\frac{g}{3}\phi_{N+1}^3(t)
\nonumber \\
&&+A_{N+1}(t)~[\phi_{N+1}(t),\dot{\phi}_{N+1}(t)]
+\frac{1}{2}~[A_{N+1}(t),\phi_{N+1}(t)]^2
+\frac{A_{N+1}^2}{\alpha}\Big\}\Big] \,.
\nonumber \\
&& \label{expAZ1_{N+1}} \ea The
$A_{N+1}(t)[\phi_{N+1}(t),\dot\phi_{N+1}(t)]$ term above is crucial
to study the nonsinglets. Even though they are present, the gauge
invariance tries to project the system to the singlet sector while
the gauge breaking term prevents to do so. In \cite{GK1}, a finitely
large radius representation of singlet sector was obtained by
throwing this particular term by hand as the nonsinglets are
confined at small temperature. In \cite{BoulKaza}, the partition
function for one vortex/anti-vortex pair, {\it i.e.} in the adjoint
representation was calculated by analytical continuation from the
twisted partition function of the standard harmonic oscillator to
that of the upside down oscillator. For $\alpha =0$, the gauge
fields are forced to vanish and the partition function reduces to
that of ungauged matrix quantum mechanics on circle.

Because of the gauge breaking term, the integration over all
possible configurations of $A(t)$ formally inserts (the gauge
invariant) operator $\exp \mbox{Tr}(-\alpha J^2)$ in the partition
function, \be \int
dA~\exp\mbox{Tr}\Big(-\frac{A^2}{\alpha}+2iAJ\Big) ~\sim~ \exp (-
\alpha J^2) ~\sim~ \exp (-N\alpha n)\,. \ee Here $J^2$ is
proportional to the quadratic Casimir invariant $C(n) \approx Nn$.
Characterizing the irreducible representations  in terms of the
number of the white boxes $n$ in the Young tableaux, the quadratic
Casimir only depends on $n$ to the leading order in $N$. The reason
for this behavior is that, $\exp \mbox{Tr}(-\alpha J^2)$ acts on the
states $\vert\mbox{Adj}\rangle$ in the adjoint representation
(belonging to the nonsinglet sector) of the MQM in the gauge
invariant way,
 \be A\cdot J\vert
\mbox{Adj}\rangle^n = \alpha~ n \vert \mbox{Adj}\rangle^n\,. \ee The
parameter $\alpha$ behaves like the fugacity of vortices. The
operator $\exp \mbox{Tr} (-\alpha J^2)$ therefore counts the vortex
number.

We will not go into the details of the RG calculation involving
Feynman diagrammatics. See the original work \cite{dd-nonsinglet}for
a complete account of the scheme involving the modified partition
function (\ref{AZ1N}). The list of required beta functions involves
in addition the one for the fugacity parameter $\alpha$. Just recall
from \cite{dd-nonsinglet} that, as in standard Wilsonian RG method,
the rescaling of the variables $t$ and the conjugate momentum
$R^{-1}$ and the fields $\phi(t)$ and $A(t)$, after evaluating the
Feynman diagrams, are performed in order to restore the original
cut-off in the following way:

\ba && t \to t'(1+h~dl)\,,~~~~~R \to R'(1+h~dl)\,,
\nonumber\\
&&\phi(t) \to \rho\phi'(t')\,,
\nonumber\\
&&A(t) \to (1-h~dl)~\eta A'(t')\,, \label{rescaling} \ea where, \be
dl=1/N\,,~~~~h = h(R) + \sum_{i,j} c_{ij}~g^i\alpha^jh_{ij}(R) \,.
\ee So the parameter $h$ is a function of the radius $R$ and the
matrix coupling $g$ and its functional form can be explicitly
determined from the behavior of the flow equations near the fixed
points. In fact $h$ turns out to be the scaling dimension of the
operator coupled with the mass parameter, {\it i.e.} the coefficient
of the $\phi(t)^2/2$ term, and appears in the universal term of the
beta function equation of the mass parameter. Being in the universal
term of the beta function equations of the couplings $g$ and
$\alpha$, the function $h$ determines the radius at which the
corresponding operators become relevant and could trigger phase
transition. Like in the previous case of ungauged model discussed in
the beginning of this section, the phase transition occurs at the
self-dual radius as the free energy changes sign due to a contest
between the entropy of the liberated vortices and the energy of the
system \cite{GK2}.

In a range of values below the self-dual radius, the pair of fixed
points become purely repulsive fixed points of large coupling and
exhibit {\it negative specific heat} and one loop correction to the
Hagedorn density of states very similar to those exhibited by an
unstable Euclidean black hole in flat space time. We will now
amplify this observation of \cite{dd-nonsinglet} in the discussion
below.

The change of entropy exhibits a discontinuity at $R=0.73$, little
above the BKT temperature, indicating the Hagedorn transition to be
of first order. Around this region the involved scaling dimensions
including $h$ are large constants. As we proceed we see black hole
like behavior emerging from the region where $h$ is a large positive
constant. Thus around $R_H$ the free energy can be written as

\be \mc{F}_s \sim f[(R/R_H-1)^{1/h}, \Delta, \hat \alpha, N]\,, \ee
where the inverse temperatures are defined by,

\be \beta = 2\pi R = \frac{1}{T}\,,~~~~\beta_H = 2\pi R_H =
\frac{1}{T_H}\,, \ee and

\be \Delta = 1-g/g^* ~~\mbox{and}~~ \hat\alpha = 1-\alpha/\alpha^*
\ee are the renormalized bulk cosmological constant and the
renormalized fugacity of the vortices respectively.

The relevant thermodynamic quantities are given by

\ba \mc{F}_s (\beta-\beta_H) &=& -\beta~F(\beta-\beta_H) = \ln
Z(\beta-\beta_H) \,,
\nonumber \\
E &=& \frac{\partial(\beta F)}{\partial\beta} = \frac{-\partial \ln
Z}{\partial\beta} \,,
\nonumber \\
C_v &=&-\beta^2 \Big(\frac{\partial E}{\partial \beta}\Big)_v\,. \ea
Using $\mc{F}_s$, we have

\ba E &\sim& \frac{1}{h}~\beta_H^{-1}~\Big(\frac{\beta-\beta_H}
{\beta_H}\Big)^{1/h-1}\,.
\nonumber \\
\label{energy} \ea Since near the phase transition the fluctuation
of energy is large, the canonical ensemble would diverge. In such
situation, it is better to pass to the microcanonical ensemble with
fixed energy and the temperature defined by

\be \beta = \frac{\partial S(E)}{\partial E} \,. \label{microtemp}
\ee Using (\ref{energy}) for large positive $h$, one can solve for
$\beta$ in terms of $E$ as \be \beta-\beta_H \sim
-\frac{h^{-1}}{E}\,. \label{bh-entropy} \ee Combining this with the
definition of temperature in the microcanonical ensemble
(\ref{microtemp}) one can calculate the near Hagedorn one loop
finite energy  correction to the usual definition of Hagedorn
density of states, $\rho (E) = \exp[S(E)] \sim \exp[\beta_H E]$, and
the usual inverse temperature, which is otherwise a constant
$\partial S/\partial E = \beta_H$. The finite energy corrections are
of the form

\ba \beta &=& \frac{\partial\ln\rho}{\partial E} = \beta_H +
\frac{s_1}{E}  + O\bigg(\frac{1}{E^2}\bigg) \,,
\nonumber \\
\rho(E) &\sim& E^{s_1}~\exp[\beta_H
E]\bigg[1+O\bigg(\frac{1}{E}\bigg)\bigg] \,. \label{oneloopcorr} \ea
Here the number $s_1$  comes from the one loop correction. If $s_1$
is negative, the specific heat is negative, {\it i.e.} increasing
the energy of the system gives rise to the decrease of temperature,
indicating {\it Euclidean black hole like behavior in flat
spacetime}.

Using (\ref{microtemp}) and (\ref{energy}) the one loop correction
is identified as follows:

\ba S(E) \sim \beta_H E-\frac{1}{h} \ln E\,,
\nonumber\\
\rho(E) \sim E^{-\frac{1}{h}} \exp [\beta_H E]\,,~~~s_1=-1/h < 0\,.
\label{entropy-dos} \ea In fact this behavior is true for large
positive $h$ and the corresponding range of radius only. As
mentioned above, the RG analysis in \cite{dd-nonsinglet} shows that
such fixed points of very large positive $h$ in the region below the
self-dual radius indeed exist. These are purely repulsive fixed
points of large (diverging) coupling $g^*\to \infty,\, \alpha^*=0$
and with positive string susceptibility exponent $\gamma_0\sim 2$.
The specific heat is negative and is given by

\be C_v \sim -\frac{1}{h}~\frac{\beta^2}{(\beta-\beta_H)^{2}} \,.
\label{sp-heat} \ee Hence we call the fixed points as the {\it
Euclidean black hole fixed points}. Using the relation $C_v = -\beta
(\partial S/\partial\beta)_v$ we see discontinuity in entropy (a
measure of the latent heat of the transition),

\be S(\beta-\beta_H) \sim -\frac{1}{h}\Big(\log (\beta-\beta_H)-
\frac{\beta_H}{\beta-\beta_H}\Big)\,, \label{entropy-thermal} \ee
suggesting that the Hagedorn transition is a first order phase
transition at little higher temperature than the BKT temperature,
driving the system to an unstable (and possibly a black hole) phase.

To summarize, $h$ determines the scaling exponents of the fixed
point by saturating itself to a constant value characteristic to
that particular fixed point. Thus the rescaling of the
compactification radius $R$ in (\ref{rescaling}) in some sense tells
us that, as the system flows to various fixed points, the target
space geometry changes accordingly. This is something new in matrix
model, which directly enables us to determine the target space
metric around a fixed point by solving the equation in its
neighborhood

\be \frac{dR}{dl} = -hR\,. \ee For example, $h=0$ corresponds to a
$c=1$ fixed point \cite{dd-singlet} giving a flat metric. In the
rest of the paper, we will use this rescaling relation of the
compactification radius with respect to the Liouville field, acting
as the RG scale, to extract the two dimensional Euclidean black hole
metric around the new type of fixed points arising above the BKT
transition point discussed above.

\section{The Black Hole Metric from the $R$ Trajectory}
\setcounter{equation}{0}

In this section we will show how the $2D$ black hole metric can be
directly obtained from the RG flow of the compactification radius
$R$ in the large $N$ renormalization analysis of the boundary theory
\cite{dd-nonsinglet}. Recall that the rescaling of the radius $R$,
$R\to R'(1+h~dl)$, to restore the original cut-off is in some sense
a running of the compactification radius given by the following beta
function \cite{dd-nonsinglet}:

\be \beta_R = \frac{dR}{dl} = -h R \,. \ee This indicates a
deformation of the target space geometry. From the definition of the
double scaling limit, the RG scale $l=1/N$ is given by

\be \frac{1}{N (g-g_c)^{(\gamma_0-2)/2}}=\mbox{const}\,. \ee The
constant in the right hand side is fixed by the closed string
coupling $g_s$. Near the black hole fixed point, the string
susceptibility exponent $\gamma_0=2$ \cite{dd-nonsinglet}. It is
then natural to identify the RG scale with the dilaton. This is
similar to the case of holographic RG in AdS/CFT where the beta
functions $\beta_i = d\lambda_i/d\phi$ describe the running of the
boundary couplings with respect to a RG scale $\phi$ that is
actually the scale factor of the 5D supergravity metric. Thus the
beta function equation for the compactification radius becomes

\be \frac{dR}{d\phi} = -h R \,.\ee Now, as we have seen, $h \to 0$
for the usual $c=1$ fixed point that describes flat metric with a
linear dilaton background \cite{dd-singlet}. As a result the
corresponding asymptotic radius is independent of the scale. On the
other hand, $h$ saturates to a large positive value as the radius
gets very small and the theory flows to the Hagedorn point $\beta
\to \beta_H$ \cite{dd-nonsinglet}. Hence $h$ can be taken as
independent of $R$ around the fixed point where we see black hole
like behavior with emergence of negative specific heat. Hence the
solution to the above equation associated to the region with
negative specific heat is

\be R^2 = R_0^2~e^{2h\phi_0}e^{-2h\phi} \,,\label{RflowSoln}\ee
obtained by imposing the boundary condition $R(\phi_0)=R_0$. In
Lorentz signature, the radius is given by $R^2 =
-(x-x_0)^2+(y-y_0)^2$, which can be written in terms of the light
cone coordinates $u=x+y$, $v=x-y$ as \be R^2 = -uv+(x_0-y_0)u
+(x_0+y_0)v - (x_0^2-y_0^2)\,. \ee Shifting $u$ and $v$ by
constants, this is expressed as

\be R^2 = -uv -A \,. \ee Using (\ref{RflowSoln}), the above
expression gives the solution for $\phi$ in terms of the light cone
coordinates \be e^{2h\phi} = -\frac{R_0^2~e^{2h\phi_0}}{uv+A}\,.\ee
This gives the bulk metric in the conformal gauge as follows \be
ds^2 = -e^{2\phi} (dx^2-dy^2) = -e^{2\phi}~du dv
=\frac{R_0^{2/h}~e^{2\phi_0}}{A^{1/h}(1+uv/A)^{1/h}} du dv\,.\ee

According to the results in \cite{dd-nonsinglet} and the review in
the previous section, the theory has negative specific heat
indicating black hole like behavior occurring at $\beta_H=2\pi R_H$,
where $R_H$ is just below the self dual radius (or the BKT radius).
Recall that by solving the Callan-Symanzik equation around the
black-hole like fixed point, where $h$ blows up to a large positive
constant, and the other scaling dimensions $\Omega_1 \to \infty,
~\Omega_2 \to \infty, ~\gamma_0 \to 2$\footnote{Here the quantities
$\Omega_1$ and $\Omega_2$ are scaling dimensions corresponding to
the matrix model coupling and fugacity respectively.}, the singular
part of free energy exhibits a scaling behavior

\be \mc{F}_s \sim f\Big((R/R_H-1)^{1/h}, \Delta, \hat \alpha, N
\Big)\,. \ee Here $\Delta$ and $\hat \alpha$ are the scaling
variables for the coupling and fugacity in the corresponding matrix
quantum mechanics and $N$ is the size of the matrix. As a result the
entropy and the density of state exhibit one loop correction at
large $h$, and for positive $h$, the specific heat is negative: \ba
S(E) &\sim& \beta E - \frac{1}{h}\ln E \,, \nonumber
\\ \rho (E) &\sim & E^{-1/h}\exp [\beta_H E]\,, ~~ s_1 = -1/h < 0
\,, \nonumber \\ C_v &\sim & -\frac{1}{h}~\frac{\beta^2}{(\beta -
\beta_H)^2} \,.\ea Hence for large positive $h$, we have the
following two dimensional metric \be ds^2 \simeq \frac{h
R_0^{2/h}~e^{2\phi_0}}{A^{1/h-1}} ~\frac{du dv}{uv+hA}\,.\ee This is
nothing but the two dimensional black hole metric in
Kruskal-Szekeres coordinates \cite{MSW,Witten}. The horizon is at
$uv = 0$.

\section{Holographic RG and the origin of the $R$ trajectory}
\setcounter{equation}{0}

In this section we will discuss a possible explanation of the origin
of the $R$ trajectory in the large $N$ RG from a holographic RG
perspective \cite{dvv,vv,ev,kv,db}. Here the matrix quantum
mechanics acts as the boundary theory and the two dimensional
gravity theory as the holographic dual to the boundary theory. The
boundary equations in large $N$ renormalization group flows in the
matrix quantum mechanics, derived in
\cite{dd-singlet,dd-nonsinglet}, can be thought of as some kind of
radial evolution of the scalar fields coupled to the two dimensional
gravity. In particular, from the holographic RG point of view it can
be shown that the simple flow equation for $R$ around the black hole
fixed point, can be retrieved from the ratio of flow velocities (the
radial equations of motion) of the bulk scalar fields coupled to
$2D$ dilatonic black hole background. In other words, the $2D$ cigar
black hole can be seen as the {\it a specific RG flow trajectory}
towards nontrivial IR fixed points as seen by the boundary couplings
propagating in the radial direction as the bulk scalar fields of the
dual theory~\footnote{In the context of holographic RG in ADS/CFT, a
crude analogy may be drawn to the deviation of the $5d$ geometry of
$ADS_5$ from its most symmetric form due to the evolution of the
boundary couplings in the radial direction as the scalars of the
dual supergravity. The deviated geometry may contain domain wall
structures or naked singularity as the specific RG trajectories
towards nontrivial IR QFTs \cite{GPPZ1,GPPZ2,FGPW,PS}.}.

\subsection{Classical closed string field evolution}

To evaluate the boundary RG flow for our $2D$ theory let us consider
the scalar $\phi^I=t$  (the Euclidean time) coupled to gravity and
the metric describing the space to be that of a cigar geometry

\be ds^2=k(d\phi^2+\tanh^2 \phi~ dt^2)\,. \label{cigarmetric}\ee Let
us consider a RG scale $\mu$ corresponding to the position
$\phi=\phi_0$ in the radial direction, that separates high and low
energy contributions (into local and nonlocal parts) in the
effective action. A shift in the adjustable parameter $\phi_0$
relates to the physical RG scale transformation through the shape of
the boundary metric

\be g_{\mu \nu}=a^2 (\phi) \hat g_{\mu \nu}\,,~~~~a=\mu/M_s\,.
\label{metric-rescaling} \ee The local and the nonlocal parts in the
effective action evolves under a shift in $\phi_0$ in such a way
that a unique~\footnote{The uniqueness was argued in \cite{vv}.}
classical trajectory (the field configuration on the boundary)
solves the equation of motion for the total action. The flow
velocities of the scalars with respect to the sliding scale $\phi_0$
are proportional to the variation of the local (or equivalently the
nonlocal) part of the action. The variation of the local and the non
local part together solves the classical equation of motion. Using
this in the classical Hamiltonian constraint, one gets a nonlinear
Hamilton-Jacobi evolution for the local (or equivalently the
nonlocal) part of the action which basically serves as the
Callan-Symanzik equation for Holographic RG. Let us here briefly
mention its relation to the Callan-Symanzik equation for the one
point function of the loop operator in the large $N$ RG, namely the
WdW equation, which is basically a quantized version of the
hamiltonian constraint on the world sheet. The WdW equation from the
large $N$ RG being computed by a matrix quantum mechanics, is
$\alpha'$ exact and gives a richer structure than the minisuperspace
WdW when computed with $N^2$ quantum mechanical degrees of freedom
\cite{d-infl}. On the other hand the Callan-Symanzik equation for
the Holographic RG, on being linearized by a variation with respect
to the local (or the nonlocal) part of the action, reduces to a
minisuperspace WdW with a leading order in the $\alpha'$ dependent
term.

In the following few paragraphs, we will briefly sketch few
essential points of holographic RG for finite $\alpha'$ from
\cite{kv}, which will be used in the remaining part of the paper.
Following \cite{kv}, the total low energy effective action for small
't Hooft coupling ($\lambda=N g_{YM}^2$) can be schematically
written as sum over all $n$-loop planar open string diagrams in the
closed string background $\phi$

\be S(\phi)=\Gamma_{0}(\phi)+\sum_{n \ge 1} \lambda^n \Gamma_{n}
(\phi)\,. \label{totalS-finite-a'}\ee Here $\Gamma_n (\phi)$
represent the $(n-1)$ loop open string contribution given by the
partition function of the world sheet sigma model in background
given by $\phi$ on a sphere with $n$ holes with all moduli
parameterizing the sizes and relative locations of the holes being
integrated over. $\Gamma_0$ gets contribution from sphere without
holes and can be interpreted to have the same form as the standard
classical action of a closed string field theory \cite{Zw}. This is
a finite $\alpha'$ generalization of the low energy effective action
used in the holographic set up in the context of $AdS_5/CFT_4$
duality \cite{dvv,ev,db} or in warped compactifications \cite{vv}
involving dynamical gravity.

Let us now talk about the issue of scale transformation in gravity
with respect to the action (\ref{totalS-finite-a'}). As a standard
procedure, one introduces a cut-off $a$ to regulate both the
divergences from the sigma model expectation value and that from the
integral over the moduli (when there are holes approaching each
other) leading to sigma model Weyl anomaly. The renormalized
background $\phi(a)$ with sigma model beta functions

\be a \frac{\partial \phi^I}{\partial a}=\beta_{ws}(\phi)^I \ee then
compensates the Weyl anomaly by Fischler-Susskind mechanism
\cite{fs, fs1} canceling the net cut-off dependence of the total
action $S(\phi(a),a)$

\be a \frac{d }{d a}S(\phi(a),a)= \beta^I_{ws}(\phi) \frac{\partial
S}{\partial \phi^I}+a \frac{\partial S}{\partial
a}=0\,.\label{confinv}\ee However,  for $\lambda \to 0$, this
cancelation requirement is essentially the usual condition for
conformal invariance as the holes due to open string loops are
absent.

Now the explicit scale dependence of the total action $S(\phi(a),a)$
comes from the boundary of the moduli space described by the
degenerate geometries. As pointed out in \cite{pol-projects}, in a
theory with gravity the question of scale dependence or the beta
function is solely determined by degenerate or pinched surfaces
(assuming the overall geometry to be spherical). This is because,
all the components of the energy momentum tensor, the generator of
the scale transformations, vanish or become $BRST$ commutators on
gauge fixing. Thus only the degenerate geometries, that form the
boundary of the moduli space, have a nonzero contribution to the
scale dependence.

To evaluate the Weyl anomaly due to the explicit scale dependence of
the total action, let us consider the $UV$ regulator $a$ on the
world sheet giving a lower bound to the minimal geodesic length
$l_C$ of all non-contractible contours $C$ surrounding a nonzero
number of holes, making the boundary of the regulated moduli space
to be degenerate surfaces satisfying the bound for one or more
contours $C$ . Such a pinched surface is conformally equivalent to
spheres with vanishing holes separated by long cylinder of length
$1/a$ for which the closed string propagator in the dual channel has
acquired a large length. Cutting the cylinder and inserting a
complete set of states, the partition function factorizes into a sum
of products of two one-point functions defined on each half of the
surface on each side of the long propagator. More specifically the
evolution operator along the long tube takes the form

\be a^{L_0+\bar L_0}=\frac{1}{2} \sum_{I,J} |O_I \rangle G^{IJ}
\langle O_J |\,. \label{propagator}\ee This determines the anomaly
to be of the form

\be a \frac{\partial S}{\partial a}=-\frac{1}{2} G^{IJ}
\frac{\partial S}{\partial \phi^I} \frac{\partial S}{\partial
\phi^J}\,. \label{esd}\ee Using the form of the total action
(\ref{totalS-finite-a'}), one can see the anomaly (\ref{esd}) is
only compensated by the following running of the renormalized
background $\phi^I(a)$,

\be \beta^I_{ws} (\phi)=G^{IJ} \frac{\partial \Gamma_0}{\partial
\phi^J} \,, \label{betafn-gamma0} \ee where $\partial
\Gamma_0/\partial \phi^I$ describe the divergences due to a sphere
without holes. Plugging (\ref{esd}) in (\ref{confinv}), one
structurally reproduces the flow equation derived by the classical
supergravity

\be a\frac{\partial}{\partial a}S(\phi(a),a)= \beta^I_{ws}
\frac{\partial S}{\partial \phi^I}-\frac{1}{2}\Big(\frac{\partial
S}{\partial \phi^I}\Big)^2= 0\,, \label{scale-inv} \ee with the
identification of $\Gamma_0$ to be the Einstein part of the
supergravity action. For a slowly varying background it only
includes the potential part, which in our $2D$ case is just the
tachyon background. Now using the flow velocities

\ba &&\dot \phi^I=\frac{1}{\sqrt{-g}}G^{IJ} \frac{\partial
\Gamma_0}{\partial \phi^J}\,,\nonumber\\&&\frac{1}{2}(\dot g_{\mu
\nu}-\dot g^{\lambda}_{\lambda} g_{\mu \nu})=\frac{1}{\sqrt {-g}}
\frac{\partial \Gamma_0}{\partial g^{\mu\nu}} \,, \ea the radial
evolution of the scalars in the bulk can be written as the
Hamilton-Jacobi equation \be \frac{1}{\sqrt
-g}\Big(\frac{1}{3}\Big(g^{\mu \nu} \frac{\delta \Gamma_0}{\delta
g^{\mu \nu}}\Big)^2-\Big(\frac{\delta \Gamma_0}{\delta g^{\mu
\nu}}\Big)^2-\frac{1}{2}\Big( \frac{\delta \Gamma_0}{\delta
\phi^I}\Big)^2 \Big)=\sqrt{-g}~\mc{L}(\phi,g)\,,\label{HJeqn}\ee
where $\mc{L}$ is the local Lagrangian density in the bulk. Note
that, in two dimensions the local action $\Gamma_0$ will only have a
local potential term $\int \sqrt{g}~T$, which via (\ref{HJeqn}) is
related to the spacetime potential $V(T)$ in the exactly same manner
as that of a closed string tachyon field. This is another way to see
that in $2D$, the nonlinear Hamilton-Jacobi equation (\ref{HJeqn})
actually solves for the $2D$ Tachyon background. In the next section
we will solve for this background. Using this background, we will
then determine the flow velocities of the scalars, which will enable
us to determine the $R$ trajectory.

\subsection{Solution for the $2D$ background}

Now plugging the rescaling (\ref{metric-rescaling}) in the radial
evolution of the scalar fields approximated by the Hamilton-Jacobi
equation (\ref{HJeqn}, a classical closed string field evolution
with Einstein part of the action replaced by $\Gamma_0$, we have

\be \Big[\frac{a^2}{12}\Big( \frac{\delta \Gamma_0}{\delta a }
\Big)^2-\frac{1}{16 a^2} \Big(\frac{\delta \Gamma_0}{\delta a }
\Big)^2-\frac{1}{2}\Big( \frac{\delta \Gamma_0}{\delta
\phi^I}\Big)^2 \Big]=a^8 \mc{L}\,. \label{HJ-gamma0}\ee One can then
solve for $\Gamma_0$ with proper initial conditions and determine
the beta function from (\ref{betafn-gamma0}). However, the main
difficulty is to write down the full space time Lagrangian $\mc{L}$
for finite $\alpha'$. Schematically, it will contain $\alpha'$
corrected full tachyon potential for the $2D$ space time, which we
can neglect as we are looking at the pure dilatonic black hole. It
will also have some general curvature term with corrections from all
orders and besides that there will be all orders of derivative terms
in fields $(t,\phi)$~. To get a crude estimation of $\Gamma_0$, let
us assume slowly varying field neglecting all derivative terms and
the leading curvature term $a^8 \Lambda$, where $\Lambda$ is the
rescaled scalar curvature. Considering the shape of the metric $a
\sim e^{\phi}$ and a trial solution of the form $\Gamma_0 \sim
e^{4\phi}~y(t,\phi)$, the equation (\ref{HJ-gamma0}) can be
rewritten as,

\be \Big(\frac{\partial y(\phi,t)}{\partial t}
\Big)^2-\Big(\frac{1}{6}-\frac{1}{8} e^{-4 \phi}\Big)
\Big(\frac{\partial y(\phi,t)}{\partial \phi} \Big)^2
-\Big(\frac{8}{3}-2 e^{-4 \phi}\Big) y(\phi,t)^2+ 2 \Lambda=0\,.\ee
For large string coupling $\phi \to \infty$, which implies a large
scale factor $a$ for $2D$ noncritical theory, the $\Lambda=0$
solution is of the form \be y(\phi,t)=\exp\Big[-\frac{4
c_1}{\sqrt{6-c_2^2}}\Big]~\exp\Big[\frac{4 (c_2
\phi+t)}{\sqrt{6-c_2^2}}\Big]\,, \ee where $c_1$ and $c_2$ are two
arbitrary constants. Now choosing $c_1=0$ and
$1/(c_2+\sqrt{6-c_2^2})=- 2 \nu/Q $,  the solution for $\Gamma_0$
takes the form of the usual $2D$ Tachyon background \be T(t,\phi)
\sim e^{-\nu t} e^{Q \phi/2}\,,\ee $Q$ being the background charge.
At large positive $\nu$, we have zero tachyon background admitting
the cigar geometry (\ref{cigarmetric}) \cite{MSW, Witten, EFR}.

\subsection{The $R$ trajectory as the ratio of flow velocities}

To get the $R$-trajectory we will now use the $2D$ background
determined above by the classical closed string field evolution and
plug it in the RG flow (\ref{betafn-gamma0}). From there, the ratio
of the pair of flow equations for the fields $(t, \phi)$ is given
by,

\be \frac{\partial t}{\partial \phi} = \frac{ \partial t/\partial
\ln a}{ \partial \phi/ \partial \ln a} =\frac{1}{c_2}=-2 \nu /Q
\,.\label{boundary-rg}\ee In the boundary theory, $t \sim t+R$. To
identify (\ref{boundary-rg}) with the boundary flow we now have to
relate the background charge $Q$ (the slope of the linear dilaton)
to the compactification radius $R$ of the target space coordinate
$t$. Such a background charge modifies the world sheet central
charge in $\phi$ to \be c_{\phi}=1+3Q^2\,. \ee Comparing the total
central charge with that of the $SL(2,\mc{R})_k/U(1)$ coset CFT
describing 2D cigar background, the linear dilaton slope is related
to the asymptotic radius of the cigar geometry $R_0$ via the level
$k$ as \be Q^2=\frac{2}{k}=\frac{1}{R_0^2}\,. \ee Thus from the
boundary theory point of view it will be natural to assume $Q \sim
\frac{1}{R}$ in the sense that a generic compactification circle of
radius $R$ in the $t$ direction will give rise to the compact
direction of the cigar geometry in the $2D$ continuum.

On the other hand, $t$ and $1/R$ being Fourier conjugate variables
on the compactification circle, a rescaling $t \to t(1+h~dl)$ in $t$
is compensated by a rescaling in $1/R$ in opposite way (keeping up
to $O(dl)$ term): $1/R \to (1-h~dl)/R$ in order to keep the product
intact. So the change in $R$ due to change in the scale will be the
same as that of $t$.

Hence the relation (\ref{boundary-rg}) leads to the same form of RG
flow at the boundary as obtained in \cite{dd-nonsinglet} by the
large $N$ renormalization group analysis \be \frac{d R}{d\phi}\sim
-2 \nu R\,. \label{bhrg}\ee Moreover, as $\nu \to +\infty$ the
tachyon field $T \to 0$, and the holographic RG analysis shows that
for a purely dilatonic black hole the factor $2 \nu$ is large and
positive in the flow equation (\ref{bhrg}), which is indeed the case
in the boundary theory as predicted by the direct large $N$
renormalization group analysis of the modified matrix quantum
mechanics described by (\ref{AZ1N}) \cite{dd-nonsinglet}.

One point is noteworthy here. The equation (\ref{bhrg}) for the
compactification radius naturally arises in the RG flow of the
boundary theory due to rescaling of the coordinates $t$, $1/R$. On
the other hand, in the bulk, we are actually studying one
dimensional trajectories of critical points with fixed rescaled
curvature $\hat R = \Lambda$ traced out by the RG flow:

\ba \beta^t &=& \frac{\partial t}{\partial \ln a} =
\frac{1}{\sqrt{-g}}\frac{\partial \Gamma_0}{\partial t} \,,
\nonumber \\ \beta^\phi &=& \frac{\partial \phi}{\partial \ln a} =
\frac{1}{\sqrt{-g}}\frac{\partial \Gamma_0}{\partial \phi} \,,\ea or
alternatively, one can look at the combined effect of $\partial
t/\partial \phi$. Since from the boundary theory point of view
$dt/dl \sim dR/dl$, where $dl \sim d\phi$. Hence in other words the
holographic RG flow actually captures the trajectory $dR/d\phi$,
that is observed to match with that of the flow in the boundary
theory. The fixed points in the RG flow of the boundary theory are
actually with respect to the coupling $g$ and $\alpha$ of the gauged
matrix model with an appropriate gauge breaking term considered in
\cite{dd-nonsinglet}. There the compactification radius $R$ acts
like a parameter. Thus the specific form of flow of $R$
(\ref{bhrg}), which is a specific RG trajectory in the holographic
RG point of view, describes $2D$ black hole near the black hole
fixed point (characterized by negative specific heat and Hagedorn
density of states) of the boundary theory given by particular fixed
point values $g^*$ and $\alpha^*$. Note that this should not be
confused with the fact that in the coset CFT description of the
cigar geometry ($\ref{cigarmetric}$), the asymptotic radius is
already fixed by IR regularity. The same radius is also fixed here
in the integration of equation (\ref{bhrg}) with proper boundary
condition. Also note that the issue of determining the
$R$-trajectory from holographic RG set up, which is in the leading
order in $\alpha'$ (using a slowly varying space time) does not
affect the computation. This is because, the $R$-trajectory being a
ratio of flow velocities is independent of the curvature term and
thus of $\alpha'$. Thus it is consistent to reproduce the
$R$-trajectory computed by $\alpha'$ exact matrix model from
holographic RG set up which is $\alpha'$ non exact.

\bigskip

\begin{flushleft}
{\bf Acknowledgments}
\end{flushleft}

We would like to thank Ofer Aharony, Michael Douglas, David Kutasov
and Massimo Porrati for discussions. Also we thank Ofer Aharony for
comments on an early draft of the paper. The work was partially
supported by Feinberg Fellowships, by the Israel-US Binational
Science Foundation, the European network HPRN-CT-2000-00122, the
German-Israeli Foundation for Scientific Research and Development,
by the ISF Centers of Excellence Program and Minerva.

\end{document}